\documentstyle[12pt,fleqn]{article}

\font\twlgot =eufm10 scaled \magstep1
\font\egtgot =eufm8
\font\sevgot =eufm7
\font\twlmsb =msbm10 scaled \magstep1
\font\egtmsb =msbm8
\font\sevmsb =msbm7

\newfam\gotfam

\textfont\gotfam\twlgot
\scriptfont\gotfam\egtgot
\scriptscriptfont\gotfam\sevgot

\newfam\msbfam
\textfont\msbfam\twlmsb
\scriptfont\msbfam\egtmsb
\scriptscriptfont\msbfam\sevmsb
\def\Bbb{\protect\pBbb}
\def\pBbb{\relax\ifmmode\expandafter\Bb\else\typeout{You cann't use
Bbb in text mode}\fi}
\def\Bb #1{{\fam\msbfam\relax#1}}

\def\thebibliography#1{\section*{References}\list
   {[\arabic{enumi}]}{\settowidth\labelwidth{#1}\leftmargin\labelwidth
     \advance\leftmargin\labelsep
     \usecounter{enumi}}
     \def\newblock{\hskip .11em plus .33em minus .07em}
     \sloppy\clubpenalty4000\widowpenalty4000
     \sfcode`\.=1000\relax}

\let\Large=\large

\def\op#1{\mathop{\fam0 #1}\limits}

\newcommand{\id}{{\rm Id\,}}

\newcommand{\beq}{\begin{equation}}
\newcommand{\eeq}{\end{equation}}
\newcommand{\ben}{\begin{eqnarray}}
\newcommand{\een}{\end{eqnarray}}
\newcommand{\be}{\begin{eqnarray*}}
\newcommand{\ee}{\end{eqnarray*}}
\newcommand{\bea}{\begin{eqalph}}
\newcommand{\eea}{\end{eqalph}}
\newcommand{\cA}{{\cal A}}

\newcommand{\cD}{{\cal D}}

\newcommand{\cE}{{\cal E}}
\newcommand{\cH}{{\cal H}}

\newcommand{\bL}{{\bf L}}
\newcommand{\bR}{{\bf R}}
\newcommand{\bC}{{\bf C}}

\newcommand{\bt}{\beta}

\newcommand{\la}{\lambda}
\newcommand{\La}{\Lambda}
\newcommand{\f}{\phi}

\newcommand{\Om}{\Omega}
\newcommand{\m}{\mu}

\newcommand{\g}{\gamma}
\newcommand{\G}{\Gamma}

\newcommand{\vt}{\vartheta}

\newcommand{\lng}{\langle}
\newcommand{\rng}{\rangle}

\newcommand{\si}{\sigma}
\newcommand{\Si}{\Sigma}
\newcommand{\w}{\wedge}
\newcommand{\wt}{\widetilde}
\newcommand{\wh}{\widehat}
\newcommand{\ol}{\overline}
\newcommand{\dr}{\partial}

\newcommand{\ot}{\otimes}

\newcounter{eqalph}
\newcounter{equationa}
\newcounter{remark}
\newcounter{example}
\newcounter{theorem}
\newcounter{proposition}
\newcounter{lemma}
\newcounter{corollary}
\newcounter{definition}
\newenvironment{eqalph}{\stepcounter{equation}
\setcounter{equationa}{\value{equation}}
\setcounter{equation}{0}

\begin{eqnarray}}{\end{eqnarray}\setcounter{equation}{\value{equationa}}}

\def\theremark{\arabic{remark}}
\def\therexample{\arabic{remark}}

\def\thedefinition{\arabic{definition}}

\newcommand{\mar}[1]{}

\begin{document}
\hbox{}

{\parindent=0pt

{\Large \bf Holonomic control operators in quantum completely
integrable Hamiltonian systems}

\bigskip

{\sc G.Sardanashvily}\footnote{{\it E-mail
address}: sard@grav.phys.msu.su}
\bigskip

{\it Department of Theoretical Physics,
Moscow State University, 117234 Moscow, Russia}
\bigskip

{\small
{\bf Abstract.}

We provide geometric quantization of a completely integrable
Hamiltonian system in the action-angle variables around an invariant torus
with respect to the angle polarization.
The carrier space of this quantization is the pre-Hilbert space of
smooth complex functions 
on the torus. A Hamiltonian of a completely integrable system in this
carrier space has a countable spectrum. If it is degenerate, its
eigenvalues are countably degenerate. 
We study nonadiabatic perturbations of this Hamiltonian by a term
depending on classical time-dependent parameters. 
It is originated by a connection on the 
parameter space, and is linear in the temporal derivatives of parameters.
One can choose it commuting with a degenerate Hamiltonian
of a completely integrable system.
Then the corresponding evolution operator acts in the eigenspaces of
this Hamiltonian, and is an operator of parallel displacement along 
a curve in a parameter space.

}
}

\section{Introduction}

By the well-known KAM theorem \cite{arn,laz}, there are perturbations 
of a Hamiltonian of a completely integrable Hamiltonian system around
an invariant torus such
that trajectories of a perturbed Hamiltonian system do not leave a
compact neighbourhood of this torus. Here, we show that one can obtain the
similar effect by means of a suitable control operator, though
trajectories of the control system need not live on tori. The
perturbed system that we construct depends on parameters. A Hamiltonian
of such a system is a sum of a dynamic Hamiltonian of the initial
completely integrable system and a perturbation term which is linear 
in the temporal derivatives of parameters. 
The key point is that integration of this term over time
through a parameter function depends only on a trajectory of this
function in a parameter space, but not on its parameterization by time. 
The corresponding evolution
operator can be treated as parallel displacement with respect to some
connection on a parameter space, and plays the role of a classical 
geometric Berry factor. One can use it as a control operator in a 
perturbed completely integrable Hamiltonian system. 
The goal is quantization of
such a system.
 
Note that, at present, geometric factor phenomena in quantum systems with
classical parameters attract special attention in connection with
holonomic quantum computation \cite{fuj,pach,zan}. This approach to
quantum computing is based on the generalization of Berry's adiabatic
phase to the non-Abelian case corresponding to adiabatically driving an
$n$-fold degenerate eigenstate of a Hamiltonian over the parameter
manifold \cite{wilcz}. Information is encoded in
this degenerate state, while 
the parameter manifold plays the role of a control parameter space.

There are different approaches to quantization of autonomous 
completely integrable
Hamiltonian systems \cite{gos,gutz}. An advantage of geometric
quantization
is that it remains equivalent under symplectic isomorphisms.
One has studied geometric quantization of an autonomous
completely integrable Hamiltonian system around an invariant torus
with respect to polarization generated by Hamiltonian vector fields of
first integrals of this system \cite{myk1,myk2}.
The problem is that the associated quantum algebra
contains functions which are not globally defined, and that
elements of its carrier space fail to be smooth
sections of the quantum bundle. We choose a different polarization.

Let $(Z,\Om)$ be a $2m$-dimensional symplectic manifold which admits $m$
smooth functions $F_k$, $k=1,\ldots,m$, which are pairwise in
involution and whose
differentials $dF_k$ are linearly independent almost everywhere. 
This is called a completely integrable Hamiltonian system. Let $M$ be
its (connected) compact manifold, i.e., all $F_k$ are constant on $M$.
By the well-known theorem \cite{arn,laz}, if $dF_k\neq 0$ on $M$, there
exists a small 
neighbourhood of $M$ which is 
isomorphic to the symplectic annulus 
\mar{z1}\beq
W=V\times T^m, \label{z1}
\eeq
where $V\subset \Bbb R^m$
is a nonempty (open, contractible) domain
and $T^m$ is an $m$-dimensional torus.
The $W$ is equipped with
the action-angle coordinates $(I_k,\f^k\,{\rm mod}\,2\pi)$.
With respect to these coordinates, the
symplectic form on $W$ reads
\mar{ci1}\beq
\Om=dI_k\w d\f^k, \label{ci1}
\eeq
and all $F_k$ are functions of action coordinates $(I_k)$ only. A
Hamiltonian on $W$ is an arbitrary analytic function $\cH$ of action
coordinates $I_k$. The corresponding Hamilton equation reads
\mar{zz0}\beq
\dot I_k=0, \qquad \dot\f^k=\dr^k\cH. \label{zz0}
\eeq

In order to geometrically quantize the symplectic manifold $(W,\Om)$,
we choose the angle 
polarization spanned by the almost-Hamiltonian vector fields $\dr^k$ 
of angle variables \cite{epr}. The associated quantum algebra $\cA$
consists of functions which are affine in action variables
$I_k$. We obtain the continuum set of its nonequivalent representations
by first order differential operators
in the separable pre-Hilbert space $\Bbb C^\infty(T^m)$ of smooth complex
functions on the torus $T^m$. This set is indexed by homomorphisms 
of the de Rham cohomology
group $H^1(T^m)$ of $T^m$ to the circle group $U(1)$, i.e., by 
collections $[\la_k]$, $k=1,\ldots,m$, of real numbers $\la_k\in [0,1)$.
In particular, the action operators read
\mar{ci7}\beq
\wh I_k=-i\dr_k -\la_k. \label{ci7}
\eeq
By virtue of the multidimensional Fourier theorem \cite{gal}, the
orthonormal basis
for $\Bbb C^\infty(T^m)$ consists of functions
\mar{ci15}\beq
\psi_{(n_r)}=\exp[i(n_r\f^r)], \qquad (n_r)=(n_1,\ldots,n_m)\in\Bbb Z^m.
\label{ci15}
\eeq
These are eigenvectors 
\be
\wh I_k\psi_{(n_r)}=(n_k-\la_k)\psi_{(n_r)} 
\ee
of the action operators (\ref{ci7}) and, consequently, of a Hamiltonian 
\mar{zz1}\beq
\wh \cH\psi_{(n_r)}=\cH(\wh I_j)\psi_{(n_r)}=\cH(n_j-\la_j)\psi_{(n_r)}.
\label{zz1} 
\eeq
In particular, if a Hamiltonian of a completely integrable system is
independent on some action variables, its eigenstates are countably
degenerate. In the framework of holonomic quantum computation, such a
degenerate state can be utilized for encoding information. The goal is to
build a holonomic control operator acting in this state. We construct it
depending on classical time-dependent parameters.

\section{Quantization of a completely integrable Hamiltonian system}

We follow the standard geometric quantization procedure
\cite{eche98,sni,wood}. 
Since the symplectic form $\Om$ (\ref{ci1}) is exact, the prequantum bundle
is a trivial complex line bundle $C\to W$. Let its trivialization
\mar{ci3}\beq
C\cong W \times \bC \label{ci3}
\eeq
hold fixed. Any other trivialization leads to
equivalent quantization of $W$.
Given the associated bundle coordinates $(I_k,\f^k,c)$, $c\in\bC$, on
$C$ (\ref{ci3}), one can treat its sections as smooth complex functions on
$W$.

The Konstant--Souriau prequantization formula associates to
each smooth real function $f\in C^\infty(W)$ on
$W$ the first order differential operator
\mar{lqq46}\beq
\wh f=-i\nabla_{\vt_f} + f \label{lqq46}
\eeq
on sections of $C$. Here
\be
\vt_f=\dr^kf\dr_k -\dr_kf\dr^k
\ee
is the Hamiltonian vector field of $f$, and
$\nabla$ is the covariant differential with respect to a
$U(1)$-principal connection $A$ on $C$ whose curvature form obeys the
prequantization condition $R=i\Om$. Such a connection reads
\mar{ci20}\beq
A=A_0 +icI_kd\f^k\ot\dr_c, \label{ci20}
\eeq
where $A_0$ is a flat $U(1)$-principal connection on $C$. The classes
of gauge conjugated flat
principal connections on $C$ are indexed by the set $\Bbb R^m/\Bbb Z^m$ of
homomorphisms of the de Rham cohomology  group  
\be  
H^1(W)=H^1(T^m)=\Bbb R^m  
\ee  
of the annulus $W$ (\ref{z1}) to
$U(1)$ \cite{eche98}.  Let us choose the representatives   
\be
A_0[\la_k]=dI_k\ot\dr^k + d\f^k\ot(\dr_k +i\la_kc\dr_c),   \qquad
\la_k\in [0,1),
\ee  
of these classes.
Then the connection $A$
(\ref{ci20}) up to gauge conjugation reads  
\mar{ci14}\beq
A[\la_k]=dI_k\ot\dr^k + d\f^k\ot(\dr_k +i(I_k+\la_k)c\dr_c).
\label{ci14}  
\eeq  
For the sake of simplicity, let 
$\la_k$  in the expression (\ref{ci14}) be arbitrary real numbers, but
we will bear in mind that  connections $A[\la_k]$ and $A[\la'_k]$ with
$\la_k-\la'_k\in\Bbb Z$  are gauge conjugated.  

Given a connection
(\ref{ci14}),  the prequantization operators (\ref{lqq46}) read
\mar{ci4}\beq  
\wh f=-i\vt_f +(f-(I_k+\la_k)\dr^kf). \label{ci4}  
\eeq
Since 
the divergence of any Hamiltonian vector field with respect to canonical
coordinates  vanishes, the prequantization operators $\wh f$  also
keep their form (\ref{ci4}) on sections of the quantum bundle
$C\ot\cD_{1/2}$, where $\cD_{1/2}\to W$ is a metalinear bundle, whose
sections are half-forms on $W$.  

Let us choose the manifested angle polarization. It is the 
vertical tangent bundle $V\pi$ of the fiber bundle 
\be
\pi:V\times T^m\to T^m
\ee
spanned by the vectors $\dr^k$. The corresponding quantum algebra
$\cA\subset C^\infty(W)$  consists of affine functions  
\mar{ci13}\beq
f=a^k(\f^r)I_k + b(\f^r) \label{ci13} 
\eeq 
of action coordinates $I_k$.
The carrier space $\cE$ of its representation (\ref{ci4}) consists
of sections $\rho$ of the quantum bundle $C\ot\cD_{1/2}\to W$ of
compact support which obey the condition $\nabla_\vt\rho=0$ for any
Hamiltonian vector field $\vt$ subordinate to the polarization $V\pi$.
This condition reads  
\be 
\dr_kf\dr^k\rho=0, \qquad \forall f\in
C^\infty(W). 
\ee 
It follows that elements of $\cE$ are independent of
action variables and, consequently, fail to be of compact support,
unless $\rho=0$, i.e., $\cE$ reduces to $\rho=0$.   

Therefore, let us modify the standard quantization procedure as
follows \cite{sard02}. Given an imbedding  
\be 
i_T:T^m\to V\times T^m, 
\ee 
let
$C_T=i^*_TC$ be the pull-back of the prequantum bundle $C$ (\ref{ci3})
onto the torus $T^m$. It is a trivial complex line bundle
$C_T=T^m\times\bC$ whose sections are smooth complex functions on  $T^m$.
Let   
\be  
\ol A[\la_k] = i^*_TA[\la_k]= d\f^k\ot(\dr_k +i(I_k+\la_k)c\dr_c)   
\ee
be the pull-back of the connection $A[\la_k]$ (\ref{ci14}) onto $C_T$, and 
let $\ol\nabla$ denote the corresponding covariant differential.  Let
$\cD_T$ be a 
metalinear bundle  of complex half-forms on the torus $T^m$.   It
admits the canonical lift   of any vector field $\tau$ on $T^m$, and
the corresponding  Lie derivative of its sections reads  
\be
\bL_\tau=\tau^k\dr_k+\frac12\dr_k\tau^k.   
\ee  
Let us consider the
tensor product  
\mar{ci6}\beq  
Y=C_T\ot\cD_T\to T^m. \label{ci6}  
\eeq
Since the Hamiltonian vector fields   
\be  
\vt_f=a^k\dr_k-(I_r\dr_ka^r
+\dr_kb)\dr^k  
\ee   
of functions $f$ (\ref{ci13}) are projectable onto
$T^m$, one can  associate to each $f\in\cA$ the first order
differential operator 
\mar{lmp135}\beq 
\wh f=(-i\ol\nabla_{\pi\vt_f}
+f)\ot\id+\id\ot\bL_{\pi \vt_f}=
-ia^k\dr_k-\frac{i}{2}\dr_ka^k-a^k\la_k +b \label{lmp135} 
\eeq 
on
sections of $Y$. A direct computation shows that the operators
(\ref{lmp135}) obey the Dirac condition  
\be
[\wh f,\wh
f']=-i\wh{\{f,f'\}}.  
\ee 
Sections $s$ of $Y\to T^m$
constitute a pre-Hilbert space $\cE_T$ with 
respect to the nondegenerate Hermitian form  
\be  
\lng s|s'\rng=\left(\frac1{2\pi}\right)^m\op\int_{T^m}  s \ol s', 
\qquad s,s'\in \cE_T.  
\ee  
Then $\wh f$ (\ref{lmp135}) are Hermitian
operators in $\cE_T$. They provide the desired geometric quantization of
a completely integrable Hamiltonian system on the annulus $W$
(\ref{z1}).    

Of course, this quantization depends on the choice
of a   connection $A[\la_k]$ (\ref{ci14}) and  a metalinear bundle
$\cD_T$, which need not be trivial.     

If  $\cD_T$ is trivial,
sections of the quantum bundle $Y\to T^m$ (\ref{ci6})  obey the
transformation  rule   
\be  
s(\f^k+2\pi)=s(\f^k)  
\ee  
for all indices
$k$. They are naturally complex smooth functions on $T^m$, i.e., the
carrier space $\cE_T$ coincides with the pre-Hilbert space $\Bbb
C^\infty(T^m)$ of smooth complex functions on $T^m$. Its orthonormal
basis consists of the functions $\psi_{(n_r)}$ (\ref{ci15}). The action
operators $\wh I_k$ (\ref{lmp135}) in this space take the form
(\ref{ci7}). Other elements of the algebra $\cA$ are 
decomposed  into the pull-back functions $\pi^*\psi_{(n_r)}$ on $W$ which act
on $\Bbb C^\infty(T^m)$ by multiplications  
\mar{ci11}\beq
\pi^*\psi_{(n_r)} \psi_{(n'_r)}=\psi_{(n_r)} 
\psi_{(n'_r)}=\psi_{(n_r+n'_r)}. \label{ci11}   
\eeq    

If $\cD_T$ is a nontrivial metalinear bundle, sections of the quantum
bundle  $Y\to T^m$ (\ref{ci6}) obey the transformation  rule
\mar{ci8}\beq  
\rho_T(\f^j+2\pi)=-\rho_T(\f^j)  \label{ci8}  
\eeq  
for
some indices $j$. In this case, the orthonormal basis for the 
pre-Hilbert space  $\cE_T$ can be represented by double-valued complex
functions 
\mar{ci10}\beq 
\psi_{(n_i,n_j)}=\exp[i(n_i\f^i+
(n_j+\frac12)\f^j)] \label{ci10} 
\eeq 
on $T^m$. They are eigenvectors
\be 
\wh I_i\psi_{(n_i,n_j)}=(n_i-\la_i)\psi_{(n_i,n_j)}, \qquad \wh
I_j\psi_{(n_i,n_j)}=(n_j-\la_j +\frac12)\psi_{(n_i,n_j)}  
\ee 
of the
operators $\wh I_k$ (\ref{ci7}), while the pull-back functions
$\pi^*\psi_{(n_r)}$ act on the basis (\ref{ci10}) by the above law
(\ref{ci11}). It follows that the representation of the quantum algebra
$\cA$ determined by the connection $A[\la_k]$ (\ref{ci14}) in the
space of sections (\ref{ci8}) of a nontrivial quantum bundle $Y$
(\ref{ci6})   is equivalent to its representation determined by the
connection $A[\la_i,\la_j-\frac12]$ in the space $\Bbb C^\infty(T^m)$ of
smooth complex functions on $T^m$.  Therefore, one can restrict the
study of representations of the quantum algebra $\cA$ to its
representations in $\bC^\infty(T^m)$ defined by different
connections (\ref{ci14}). These representations are nonequivalent,
unless  $\la_k-\la'_k\in\Bbb Z$ for all indices $k$.  

Given the
representation (\ref{lmp135}) of the quantum algebra $\cA$ in
$\bC^\infty(T^m)$,  any polynomial Hamiltonian $\cH(I_k)$ of a
completely integrable system is uniquely quantized as a Hermitian
element  $\wh\cH(I_k)=\cH(\wh I_k)$ of the enveloping algebra $\ol\cA$
of $\cA$. It has the countable spectrum (\ref{zz1}).
Note that, since $I_k$ are diagonal operators, one can also quantize the
Hamiltonians which are analytic functions on $\Bbb R^m$.

\section{The classical controllable completely integrable system}

A generic momentum phase space of a Hamiltonian system
with time-dependent parameters is a composite fiber bundle
\be
\Pi\to\Si\to\Bbb R,
\ee
where $\Pi\to\Si$ is a symplectic bundle, $\Si\to\Bbb R$ is a parameter
bundle whose sections are parameter functions, and $\Bbb R$ is the time
axis \cite{epr1,book98,jmp00}. Here, we assume that all bundles are
trivial and, moreover, their trivializations hold fixed. Then the
momentum phase space of a completely integrable Hamiltonian system on
the symplectic annulus $W$ (\ref{z1}) is the product
\mar{zz2}\beq
\Pi=\Bbb R\times S\times W, \label{zz2}
\eeq
equipped with the coordinates $(t,\si^\la, I_k,\f^k)$. It is convenient
to suppose for a time that parameters are dynamic variables. The
momentum phase space of such a system is
\mar{zz3}\beq
\Pi'=\Bbb R\times T^*S\times W, \label{zz3}
\eeq
coordinated by $(t,\si^\la,p_\la, I_k,\f^k)$.

The dynamics of a time-dependent mechanical system on the momentum phase
space $\Pi'$ (\ref{zz3}) is characterized by a Hamiltonian form
\mar{b4210}\beq
H_\Si= p_\la d\si^\la+ I_k d\f^k -\cH_\Si(t,\si^\m,p_\m, I_j,\f^j)dt
\label{b4210}
\eeq
\cite{book98,sard98}. For any Hamiltonian form $H_\Si$ (\ref{b4210}), there
exists a unique 
vector field $\g_H$ on $\Pi'$ such that
\be
\g_H\rfloor dt=1, \qquad \g_H\rfloor dH_\Si=0. \label{qq1}
\ee
It defines the first order differential Hamilton equation on 
$\Pi'$. 

Note that the Hamiltonian $\cH_\Si$ (\ref{b4210}) takes the form
\mar{pr30}\beq
\cH_\Si=p_\la\G^\la +I_k(\La^k +\G^\la\La^k_\la) 
+\wt\cH, \label{pr30}
\eeq
where 
\be
\La\circ\G=dt\ot(\dr_t +\G^\la\dr_\la +(\La^k + \G^\la\La^k_\la)\dr_k)
\ee
is a connection on $\Bbb R\times S\times T^m\to \Bbb R$ which is the
composition of a connection 
\mar{pr31}\beq
\G=dt\ot(\dr_t +\G^\la(t,\si^\m)\dr_\la) \label{pr31}
\eeq
on the parameter bundle $\Bbb R\times S\to\Bbb R$ and of a connection
\mar{pr32}\beq
\La=dt\ot(\dr_t+\La^k(t,\si^\m,\f^j)\dr_k) + 
d\si^\la\ot(\dr_\la +\La^k_\la(t,\si^\m,\f^j)\dr_k) \label{pr32}
\eeq
on $\Bbb R\times S\times T^m\to \Bbb R\times S$ \cite{book98,book00}. 

Bearing in mind that $\si^\la$ are parameters, one should choose the
Hamiltonian $\cH_\Si$ (\ref{pr30}) affine in momenta $p_\la$.  
Furthermore, in order to describe a Hamiltonian system with a fixed
parameter function  $\si^\la=\xi^\la(t)$, one defines the
connection
$\G$ (\ref{pr31}) such that
\be
\nabla^\G \xi=0,\qquad \G^\la(t,\xi^\m(t))=\dr_t\xi^\la. 
\ee
Then the pull-back
\be
H=\xi^*H_\Si=I_k d\f^k - (I_k[\La^k(t,\xi^\m,\f^j)+ \La^k_\la
(t,\xi^\m,\f^j)\dr_t\xi^\la]
+\wt\cH(t,\xi^\m,I_j,\f^j))dt 
\ee
is a Hamiltonian form on $\Bbb R\times W$. Let us put
\be
\wt\cH=\cH-I_k\La^k.
\ee
Then the Hamiltonian form
\mar{zz5}\beq
H=I_k d\f^k -\cH'dt=I_k d\f^k - [I_k\La^k_\la
(t,\xi^\m,\f^j)\dr_t\xi^\la+ \cH(I_j)]dt \label{zz5}
\eeq
describes a time-dependent perturbed completely integrable Hamiltonian
system on $\Bbb R\times W$. The corresponding Hamilton equation reads
\mar{zz6}\beq
\dr_t I_k=-\dr_k\La^j_\la I_j\dr_t\xi^\la, \qquad \dr_t\f^k=\dr^k\cH
+\La^k_\la \dr_t\xi^\la. \label{zz6}
\eeq

In order to make the term
\mar{zz8}\beq
\Delta=I_k\La^k_\la\dr_t\xi^\la \label{zz8}
\eeq
in the perturbed Hamiltonian $\cH'$ (\ref{zz5}) to a control operator, let us
assume that the coefficients $\La^k_\la$ of the connection (\ref{pr32})
are independent 
of time. Then, in view of the trivialization (\ref{zz2}), the
second term 
\be
d\si^\la\ot(\dr_\la +\La^k_\la(\si^\m,\f^j)\dr_k) 
\ee
of the connection (\ref{pr32}) can be seen as a connection
on the fiber bundle $S\times T^m\to S$. Let the dynamic Hamiltonian
$\cH$ be independent of action variables with some $l$ indices $a,b,c,\ldots$,
and let the perturbation term $\Delta$ (\ref{zz8})
be independent of the action and angle variables with the rest indices
$i,j,k,\ldots$. Then the Hamilton equation (\ref{zz6}) falls into the two
independent equations
\mar{zz10,1}\ben
&& \dr_t I_k=0, \qquad \dr_t \f^k=\dr^k\cH, \label{zz10}\\
&& (a)\,\,\dr_t I_a=-I_b\dr_a\La^b_\la \dr_t\xi^\la, \qquad (b)\,\, 
\dr_t\f^a=\La^a_\la \dr_t\xi^\la. \label{zz11}
\een
The first equation (\ref{zz10}) keeps the form of the Hamilton equation
(\ref{zz0}) of an 
autonomous completely integrable system, while the second one is the
control equation as follows.

Let us rewrite the equation (\ref{zz11}b) as the countable system of equations
\mar{zz12}\beq
\dr_t\psi_{(n_a)}=i\psi_{(n_a)} n_a\dr_t\f^a=
i\psi_{(n_a)}n_a\La^a_\la \dr_t\xi^\la \label{zz12}
\eeq
for functions $\psi_{(n_a)}$ (\ref{ci15}).
The left hand-side of these equations is
a multidimensional Fourier series with time-dependent coefficients. Therefore,
the equations (\ref{zz12}) for all collections of $l$ integers
$(n_a)$ make up a countable 
system of ordinary linear differential equations with time-dependent
coefficients: 
\be
\dr_t\psi_{(n_a)}= iM_{\la(n_a)}^{(n_b)}(\xi^\m)\dr_t\xi^\la \psi_{(n_b)}.
\ee
Its solution with the initial dates $\f^a(0)$ can be written formally as the
time-ordered exponential 
\mar{zz14}\ben
&&\psi_{(n_a)}(t)=U(t)_{(n_a)}^{(n_b)} \psi_{(n_b)}(0), \nonumber\\
&& U(t)=T\exp\left[i\op\int^t_0 M_\la(\xi^m(t'))\dr_t\xi^\la dt'\right]
=T\exp\left[i\op\int_{\xi([0,t])}M_\la(\si^\m)d\si^\la\right], \label{zz14}
\een
\cite{lam,oteo}. A glance at the evolution operator $U(t)$ (\ref{zz14}) shows
that solutions $\psi_{(n_a)}(t)$ of the equations (\ref{zz12}) 
are functions of a point $\si$ of the curve $\xi:\Bbb R\to S$ in the
parameter space $S$.

Substituting this solution into the equation (\ref{zz11}a), we obtain the
system of $l$ ordinary linear differential equations with time-dependent
coefficients:
\be
\dr_t I_a=-L_{\la a}^b (\psi_{(n_a)}(t),\xi^\m(t))
 I_b\dr_t\xi^\la, \qquad L_{\la a}^b=\dr_a\La^b_\la.
\ee
Its solution is given by the time-ordered
exponential 
\be
&& I_a(t)=U(t)_a^b I_b(0), \\
&& U(t)=T\exp\left[-\op\int^t_0 L_\la(\psi_{(n_a)}(t'),\xi^\m(t'))\dr_t\xi^\la
dt'\right]=
 T\exp\left[-\op\int_{\xi([0,t])}L_\la(\psi_{(n_a)}(\si),
\si^\m)d\si^\la\right]. 
\ee
This solution is also a functions of a point $\si$ of the curve $\xi:\Bbb R\to
S$ in the parameter space $S$. 

Consequently, the equation (\ref{zz11})
can be regarded a
control equation. In particular, one can choose a parameter
function $\xi(t)$ such that the trajectory of the perturbed system
does not leave a compact neighbourhood of 
the invariant torus $T^m$ of the initial completely integrable
Hamiltonian system.

\section{The quantum controllable completely integrable system}

Let us quantize the classical controllable completely integrable system
on the momentum phase space $\Bbb R\times W$ in previous Section.

The manifold 
$W'=\Bbb R\times W$,
coordinated by $(t,I_k,\f^k)$, is provided with the Poisson structure
\be
\{f,f'\}=\dr^kf\dr_kf'-\dr_kf\dr^kf', \qquad f,f'\in C^\infty(W').
\ee
This is the direct product of the symplectic structure $\Om$
(\ref{ci1}) on the symplectic annulus $W$ (\ref{z1}) and of the zero
Poisson structure on the time axis $\Bbb R$. 
In particular, the Poisson algebra
$(C^\infty(W'),\{,\})$ of 
smooth real functions on $W'$ is the Lie algebra over the ring
$C^\infty(\Bbb R)$ of smooth real functions on $\Bbb R$. 
In order to quantize the 
Poisson manifold $(W',\{,\})$, we therefore can essentially simplify
the general procedure of
instantwise 
geometric quantization of time-dependent Hamiltonian systems in Ref.
\cite{sard02}. Namely, we 
repeat geometric quantization of the symplectic manifold $(W, \Om)$ in
Section 2, but replace functions on $T^m$ with those on $\Bbb R\times
T^m$.

Let us choose the angle polarization spanned by the vectors $\dr^i$.
The corresponding quantum algebra $\cA\subset C^\infty(W')$ consists of
affine functions
\be
f=a^i(t,\f^j)I_i + b(t,\f^j) 
\ee 
of action coordinates $I_i$. It is represented by the first order
differential operators
\mar{z61}\beq 
\wh f=
-ia^i\dr_i-\frac{i}{2}\dr_ia^i-a^i\la_i +b, \qquad
\la_i\in\Bbb R, \label{z61}  
\eeq 
in the space $\Bbb C^\infty(\Bbb R\times T^m)$ of smooth complex functions on
$\Bbb R\times T^m$. Given different collections of real numbers $(\la_i)$
and $(\la'_i)$, the representations (\ref{z61}) are nonequivalent, unless 
$\la_i-\la'_i\in\Bbb Z$ for all indices $i$. The carrier space
$\Bbb C^\infty(\bR\times T^m)$ is provided with the structure of the
pre-Hilbert $\Bbb C^\infty(\Bbb R)$-module with respect to the nondegenerate 
$\Bbb C^\infty(\Bbb R)$-bilinear form
\be
\lng \psi|\psi'\rng=\left(\frac1{2\pi}\right)^m\op\int_{T^m}  \psi \ol \psi', 
\qquad \psi,\psi'\in \bC^\infty(\Bbb R\times T^m).  
\ee 
The basis for this pre-Hilbert module is made up by the pull-backs onto
$\Bbb R\times T^m$ of 
functions $\psi_{(n_r)}$ (\ref{ci15}) on $T^m$. They are the eigenvectors
of the action operators
\be
\wh I_k=-i\dr_k -\la_k, \qquad  \wh
I_k\psi_{(n_r)}=(n_k-\la_k)\psi_{(n_r)}. 
\ee

As in previous Section, let us assume that 
the dynamic Hamiltonian
$\cH$ is independent of action variables with some $l$ indices $a,b,c,\ldots$,
and let the perturbation term $\Delta$ (\ref{zz8})
be indepependent of the time and the action-angle variables with the
rest indices 
$i,j,k,\ldots$, i.e., the perturbed Hamiltonian $\cH'$ (\ref{zz5})
takes the form
\be
\cH'=\La^a_\la(\xi^\m,\f^b)\dr_t\xi^\la I_a +\cH(I_j). 
\ee
The perturbation term
\be
\Delta=\La^a_\la(\xi^\m,\f^b)\dr_t\xi^\la I_a 
\ee
of this Hamiltonian is an element of the quantum algebra $\cA$, and is
quantized by the operator
\be
\wh\Delta=-i\La^a_\bt\dr_t\xi^\bt\dr_a -\frac{i}{2}\dr_a(\La^a_\bt)
\dr_t\xi^\bt -\la_a \La^a_\bt\dr_t\xi^\bt. 
\ee
The (polynomial or analytic) dynamic Hamiltonian $\cH(I_j)$ is
quantized as in Section 2, i.e.,  $\wh\cH=\cH(\wh I_j)$. 

Since the operators $\wh\Delta$ and $\wh\cH$ mutually commute, the
corresponding quantum evolution operator reduces to the product
\mar{zz23}\beq
T\exp\left[-i\op\int_0^t\wh\cH'dt'\right]=
U_1(t)\circ U_2(t)=
T\exp\left[-i\op\int_0^t\wh\cH dt'\right]\circ 
T\exp\left[-i\op\int_0^t\wh\Delta dt'\right].
\label{zz23}
\eeq
The first operator in this product is the dynamic evolution operator of
the quantum completely integrable Hamiltonian system. It reads
\mar{zz24}\beq
U_1(t)\psi_{(n_j)}=
\exp[-i\cH(n_j-\la_j)t]\psi_{(n_j)}. \label{zz24}
\eeq
Its eigenvalues are countably degenerate.
Recall that the operator
(\ref{zz24}) acts in the $\Bbb C^\infty(\Bbb R)$-module $\Bbb
C^\infty(\Bbb R\times T^m)$. Its eigenvalues
are smooth complex functions on $\Bbb R$, and its eigenspaces
are $\Bbb C^\infty(\Bbb R)$-submodules of $\Bbb
C^\infty(\Bbb R\times T^m)$ of countable rank.

The second multiplier in the product (\ref{zz23}) is 
\mar{zz25}\ben
&& U_2(t)=T\exp\left[\op\int_0^t\{
-\La^a_\bt(\f^b,\xi^\m(t'))\dr_a -\frac12\dr_a\La^a_\bt(\f^b,\xi^\m(t'))
+i\la_a \La^a_\bt(\f^b,\xi^\m(t'))\}\dr_t\xi^\bt dt'\right]\nonumber\\
&&\qquad =T\exp\left[\op\int_{\xi([0,t])}\{
-\La^a_\bt(\f^b,\si^\m)\dr_a -\frac12\dr_a\La^a_\bt(\f^b,\si^\m)
+i\la_a \La^a_\bt(\f^b,\si^\m)\} d\si^\bt\right]. \label{zz25}
\een
It acts in the eigenspaces of the operator $U_1(t)$. For instance, 
such a space is exemplified by the pre-Hilbert $\Bbb C^\infty(\Bbb
R)$-submodule  
$\cE_0\subset \Bbb
C^\infty(\Bbb R\times T^m)$ whose orthonormal basis is made up by  
functions $\psi_{(n_a)}$ for all collections of integers $(n_a)$.
Written with respect to this basis, the operator 
$U_2(t)$ acts on $\cE_0$ as a matrix of countable rank. 

A glance on the expression (\ref{zz25}) shows that, in fact, the operator
$U_2(t)$ depends on the curve 
$\xi([0,1])\subset S$ in the parameter space $S$. One can treat it as
an operator of parallel
displacement with respect to a connection in the $\Bbb C^\infty(\Si)$-module 
of smooth complex functions on $\Si\times T^m$ along the curve
$\xi$ \cite{epr1,book00,jmp00}.
For instance, if $\xi([0,1])$ is a loop in $S$, the operator $U_2$
(\ref{zz25}) is  
the geometric Berry factor. In this case, one can think of $U_2$ as
being the holonomic control operator.

It should be emphasized that the adiabatic assumption has never been 
involved.

\end{document}